\documentclass[doublecol,linenumbers]{epl2}

\title{Spin-nematic order induced superconductivity.}
\shorttitle{Superconductivity in $FeSe$.}

\author{Naoum Karchev}

\institute{Department of Physics, University of Sofia, 1164 Sofia, Bulgaria}


\newcommand{\be}{\begin{equation}}
\newcommand{\ee}{\end{equation}}
\newcommand{\bea}{\begin{eqnarray}}
\newcommand{\beaa}{\begin{eqnarray*}}
\newcommand{\eea}{\end{eqnarray}}
\newcommand{\eeaa}{\end{eqnarray*}}

\usepackage{epsfig}
\usepackage{color}
\usepackage{amsmath}
\usepackage{amssymb}
\usepackage{bbm}
\usepackage{graphicx}

\abstract {We explore a spin-fermion model with fermion-spin-quadrupolar interaction. In a nematic phase, this interaction reduces to a four-fermion interaction that is the basis of superconductivity. When the coupling constant is positive the superconductivity is a p-wave with spin-parallel paired fermions. When it is negative the superconductivity is a p-wave and fermions are spin-antiparallel paired. For a system with zero chemical potential, even a very small coupling can bind fermions into bound state that leads to the superconductivity.
When the chemical potential is non-zero the system possesses quantum critical transition from the normal spin-nematic phase to the phase where superconductivity coexists with spin-nematicity. The value of the quantum critical fermion-spin-nematicity coupling constant depends on the chemical potential.}

\begin{document}

\maketitle

\section{Introduction}

The interplay of superconductivity and magnetism is a very old topic in solid state physics \cite{Ginzburg57}.
A common feature of magnetic superconductors is that superconductivity arises close to the magnetic quantum critical point. In the ferromagnetic superconductors $UGe_2$ \cite{UGe200} and $UThGe$ \cite{URhGe01}, the superconductivity occurs only in the ferromagnetic phase near the quantum critical point and the ferromagnetic order is stable within the superconducting phase (neutron scattering experiments). In the high-$T_c$ cuprates \cite{BM86} superconducting phase lies in the paramagnetic phase near the quantum critical point. The coexistence of the superconductivity and antiferromagnetism in the heavy fermion compounds $CeCu_2Si_2$ \cite{CeCuSi79} and $CeRhIn_5$ \cite{CeRhIn00,CeRhIn01,CeRhIn07} is more complicated.
The temperature-pressure phase diagrams show that at low pressure the systems are in antiferromagnetic state, they are superconductors in large pressure region and magnetism and superconductivity coexist in-between.

The spin systems in the nematic phase have zero magnetization. The spin-nematic order, also known as quadrupolar order, breaks down the discrete lattice  rotational symmetry but preserves time-reversal symmetry. The order parameter is an element of the tensor representation of the spin group $O(3)$\cite{Andreev84}.

Nematic order in $FeSe$ is a well-established experimental fact. At $90K$ the system undergoes a structural tetragonal-to- orthorhombic transition \cite{McQueen09} accompanied by a transition to spin nematic state. Other parents of iron-based superconductors undergo a structural transition followed closely by magnetic transition.  $FeSe$ is unusual in that no long-range magnetic order has been detected. This make the study of superconductivity very attractive. One has to investigate the interaction in FeSe which is unique for the nematic order in the material. We think that the impact of the fermion-quadrupolar interaction is what distinguishes FeSe from other iron-based superconductors. Superconductivity was found below $8K$ \cite{Hsu08}. We focus on $FeSe$ compound because its structure is very simple. The $Fe$ $3d$-electrons occupy five degenerate levels. The crystal field results in a particular splitting of the five $d$-orbitals into well separated in energy two groups: the $\emph{e}_g$ and $\emph{t}_{2g}$ states. The $\emph{t}_{2g}$ electrons form a triplet, and the $\emph{e}_g$ sectors form a doublet. In the $FeSe$ systems the ligands-$Se$ surrounding magnetic $Fe$ ions form a tetrahedral structure, thus the energy of doublet state $(d_{x^2-y^2},d_{3z^2-r^2})$ is lower. The  $\emph{e}_g$ electrons are localized and one can study them
by means of an effective spin-1 Hamiltonian with quadratic and biquadratic exchange interactions
\be \label{nematic1}
H_s\,=\,  J\sum\limits_{<ij>}\left[\cos\gamma\left( {\bf S}_i \cdot {\bf S}_{j}\right)\,+\,  \sin\gamma\left({\bf S}_i \cdot {\bf S}_{j}\right)^2\right],
\ee
where ${\bf S}_i$ are the spin operators, the sum is over all sites of a three-dimensional cubic lattice, and
$\langle i,j\rangle$ denotes the sum over the nearest neighbors. The overall exchange constant $J$ is positive and the angle $\gamma$ takes value in the interval $[0,2\pi]$.

To study quadrupolar order one introduces the quadrupolar operators \cite{Andreev84}
\be\label{nematic2}
S^{ab}_i\,=\,\frac 12 \left(S^{a}_iS^{b}_i\,+\,S^{b}_iS^{a}_i\right), \ee
where $a,b=x,y,z$.
The Hamiltonian Eq.(\ref{nematic1}) can be rewritten in the terms of spin and quadrupolar operators
\be \label{nematic4}
H_s = J\sum\limits_{<ij>}\left[(\cos\gamma-\frac 12 \sin\gamma)\left( {\bf S}_i \cdot {\bf S}_{j}\right)+\sin\gamma \left(S^{ab}_iS^{ab}_{j}\right)\right].
\ee

The spin one representation of $O(3)$ group is three dimensional. It is convenient to introduce $3\times3$ matrixes $E^{ab}$ with only one non-vanishing matrix element, at the intersection of the $a$th row and $b$th column, that is equal to unity. The $(\alpha\beta)$ matrix element of the $E^{ab}$ matrix reads
\be\label{nematic4a}
E^{ab}_{\alpha\beta} = \delta_{a\alpha}\delta_{b\beta} \ee
where $\alpha,\beta=x,y,z$.
In the terms of $E^{ab}$ matrixes the matrix elements $\alpha\beta$  of spin and quadrupolar operators have the form
\be\label{nematic9}
\hat{S}_{\alpha\beta}^a\,=\,-i\varepsilon_{abc}E^{bc}_{\alpha\beta}, \quad \hat{S}^{ab}_{\alpha\beta}\,=\,\delta_{\alpha\beta}\delta_{ab}\,-\,\frac 12 \left(E^{ab}_{\alpha\beta}\,+\,E^{ba}_{\alpha\beta}\right). \ee

The next step is to generalize the Schwinger-bosons approach. To this end we introduce three Bose operators $\varphi^a_i$ and $\varphi^{a+}_i$ \cite{Papanicolau84,Penc06}.
To separate the three dimensional Hilbert space, per lattice site, for spin one system from the infinite dimensional space of the Schwinger-bosons
we impose the operator equation
\be\label{nematic14}
\varphi^{x+}_i\varphi^x_i\,+\,\varphi^{y+}_i\varphi^y_i\,+\,\varphi^{z+}_i\varphi^z_i\,=\,\textbf{1}. \ee
The representation of the spin and quadrupolar operators by means of  Schwinger-bosons is
\bea\label{nematic13}
S^a_i & = & \varphi^{\alpha+}_i\hat{S}_{\alpha\beta}^a \varphi^\beta_i\,=\, -i\varepsilon_{abc}\varphi^{b+}_i\varphi^c_i,\nonumber \\
\\
S^{ab}_i & = &  \varphi^{\alpha+}_i\hat{S}^{ab}_{\alpha\beta}\varphi^\beta_i\,=\, \delta_{ab}\,-\,\frac 12 \left(\varphi^{a+}_r\varphi^b_r\,+\,\varphi^{b+}_r\varphi^a_r\right).\nonumber \eea
In the terms of Schwinger-bosons the spin Hamiltonian (\ref{nematic1}) reads
\bea \label{nematic17}
H_s\, & = & \,J \sum\limits_{<ij'>}\left[\cos\gamma\, \varphi^{a+}_i \varphi^a_{j}\varphi^{b+}_{j}\varphi^b_i \right. \nonumber \\
       & & \left. +(\sin\gamma-\cos\gamma)\varphi^{a+}_i\varphi^{a+}_{j}\varphi^b_{j} \varphi^b_i\right].\eea
To identify the ground state in semiclassical limit we replace the  Schwinger-boson operators by classical complex fields
$f^a_r$ and $\bar{f^a_r}$.
The energy of the system is given by the classical expression of the Hamiltonian (\ref{nematic17})
\be\label{nematic20}
H_s\, = \,J \sum\limits_{<ij>}\left[\cos\gamma\, |\bar{f^a_i} f^a_{j}|^2\, +\,(\sin\gamma-\cos\gamma)|f^a_i f^a_{j}|^2\right].
\ee
For values of the angle in the interval
$\frac 54 \pi< \gamma < \frac 32 \pi$ \cite{Papanicolau84}
the semiclassical energy (\ref{nematic20}) attains its absolute minimum if
$|\bar{f^a_r} f^a_{r'}|\,=\,1$ and $|f^a_r f^a_{r'}|\,=\,1$.
The solution of these equations which does not depend on the lattice site and satisfies the condition (\ref{nematic14}) is
$f^x=0$, $ f^y=0$ and $f^z=1$.

Replacing the Schwinger operators in equation (\ref{nematic13}) by the solution we obtain that the magnetization $<\textbf{S}>$ is zero, while the result of the explicit calculation of the quadrupolar order parameter $<S^{ab}>$ shows
that only two of the components of the tensor are no-zero
\be\label{nematic29}
<S^{xx}>\,=\,<S^{yy}>\,=\,1.\ee
The expression (\ref{nematic29}) for $<S^{ab}>$ differs from its symmetric value $\frac 23 \delta_{ab}$. Therefore, the spin rotation symmetry is broken but time reversal symmetry is preserved. This type of spin nematic state is ferroquadrupolar, since the order parameter does not depend on the
lattice site.

When the angle $\gamma$ in the Hamiltonian (\ref{nematic1}) runs the interval $\frac 12 \pi< \gamma < \frac 54 \pi $ the system is \textbf{ferromagnetic}. For values of the angle $\gamma$ in the interval $-\frac 12 \pi< \gamma < \frac 14 \pi$ the ground state of the Hamiltonian (\ref{nematic1}) is \textbf{antiferromagnetic} \cite{Papanicolau84,Penc06}.

\section{Fermion-Quadrupolar Interaction}

As mentioned above the selenium ligands surround the iron ions and form tetrahedral structure. Hence the energy of the $t_{2g}$ triplet state $(d_{xy},d_{xz},d_{yz})$ is higher than the energy of $e_{g}$ doublet. Moreover, the energy levels in the triplet are split and the $d_{xy}$ electrons occupy the lowest one. This permits to consider one band theory of electrons. With these assumptions
the Hamiltonian of the 3d electrons of iron is
\be\label{nematicSc1}
H\,=\,H_s\,+H_{f}\,+H_{fq}, \ee
where $H_s$ is the spin Hamiltonian (\ref{nematic1}), $H_f$ is the free fermion Hamiltonian
\be\label{nematicSc2}
H_f\, =\, -t\sum\limits_{< ij>} \left( c_{i\sigma }^ + c_{j\sigma } + h.c. \right) -\mu \sum\limits_{i} {n_i}, \ee
and $H_{fq}$ is the Hamiltonian of the fermion-quadrupolar interaction.

To construct the interaction we follow the idea of the Schwinger-bosons approach. To this end, we introduce a three-component spin-one vector  built by fermions in momentum space representation
\bea\label{nematicSc3}
D^x_k & = & \frac {1}{2} \left[c_{k\downarrow}c_{-k\downarrow}\,-\,c_{k\uparrow}c_{-k\uparrow}\right] \nonumber \\
D^y_k & = & \frac {1}{2i}\left[c_{k\downarrow}c_{-k\downarrow}\,+\,c_{k\uparrow}c_{-k\uparrow}\right]  \\
D^z_k & = & \frac {1}{2}\left[c_{k\uparrow}c_{-k\downarrow}\,+\,c_{k\downarrow}c_{-k\uparrow}\right], \nonumber \eea
where the wave vector $k$ runs over the first Brillouin zone of a simple cubic lattice. The components of the vector satisfy $D^a_k=-D^a_{-k}$ due to the fermion exchange. We rewrite the vector in coordinate space
\be\label{nematicSc4}
\textbf{D}_j\,=\,\frac {1}{\sqrt{N}}\sum\limits_{k} e^{i\textbf{r}_j\cdot\textbf{k}}\textbf{D}_k  \ee
and, following the prescription from equation (\ref{nematic13}),  introduce the tensor operator $T_j^{ab}$
\be\label{nematicSc5}
T_j^{ab}\,=\, D^{\alpha+}_j \hat{S}^{ab}_{\alpha\beta}D^{\beta}_j, \ee
where $\hat{S}^{ab}_{\alpha\beta}$ is the matrix representation (\ref{nematic9}) of the quadrupolar operator. By means of the spin quadrupolar operator (\ref{nematic13}) and tensor (\ref{nematicSc5}) we construct the fermion-quadrupolar interaction
\be\label{nematicSc6}
H_{fq}\,=\,\kappa \sum\limits_{j ab}T_j^{ab}S_j^{ba}\ee

In the semiclassical limit, we replace the quadrupolar operator $S^{ba}_i$ by its average value $<S^{ba}_i>$
\be\label{nematicSc7}
H_{fq}\longrightarrow\kappa\sum\limits_{j ab}T_j^{ab}<S_j^{ba}> \ee
In the nematic phase only two of the components of the tensor are not equal to zero (\ref{nematic29}) and the fermion-quadrupolar interaction reduces to the four-fermion interaction
\bea\label{nematicSc8}
H_{f^4} & = & \frac {\kappa}{2} \sum\limits_{k}\left(-c_{-k\downarrow}^{+}c_{k\downarrow}^{+}c_{k\uparrow}c_{-k\uparrow}-
c_{-k\uparrow}^{+}c_{k\uparrow}^{+}c_{k\downarrow}c_{-k\downarrow}\right. \nonumber \\
\\
& + & \left. \frac 12 \left[c^{+}_{-k\downarrow}c^{+}_{k\uparrow}+c^{+}_{-k\uparrow}c^{+}_{k\downarrow}\right]
\left[c_{k\uparrow}c_{-k\downarrow}+c_{k\downarrow}c_{-k\uparrow}\right]\right)\nonumber. \eea

\section{Superconductivity}

Following the standard procedure we present the Hamiltonian (\ref{nematicSc8}) in the Hartree-Fock approximation
\bea\label{nematicSc10}
H_{HF}& = & \sum\limits_{k}\left(\frac {1}{\kappa}\bar{\Delta}_k^{p}\Delta_k^{p}-\frac {1}{\kappa}\bar{\Delta}_k^{ap}\Delta_k^{ap} +\varepsilon_k c^+_{k\sigma}c_{k\sigma}\right. \nonumber \\
& & \left. +  \frac 12 \Delta_k^{ap} \left[c^{+}_{-k\downarrow}c^{+}_{k\uparrow}+c^{+}_{-k\uparrow}c^{+}_{k\downarrow}\right]\right. \nonumber \\
& & \left. + \frac 12 \bar{\Delta}_k^{ap}\left[c_{k\uparrow}c_{-k\downarrow}+c_{k\downarrow}c_{-k\uparrow}\right]\right. \\
& & \left. - \frac 12 \Delta_k^{p} \left[c^{+}_{-k\downarrow}c^{+}_{k\downarrow}+c^{+}_{-k\uparrow}c^{+}_{k\uparrow}\right]\right. \nonumber \\
& & \left. - \frac 12 \bar{\Delta}_k^{p}\left[c_{k\uparrow}c_{-k\uparrow}+c_{k\downarrow}c_{-k\downarrow}\right]\right),\nonumber \eea
where $\varepsilon_k = -2t\left (\cos k_x + \cos k_y + \cos k_z \right )$ and $\Delta_k^{ap}, \Delta_k^{p}$ are the gap functions.
 To obtain the representation (\ref{nematicSc10}) for the Hamiltonian $H_{HF}$ we utilized the identity
\be\label{nematicSc10a}
<c_{k\uparrow}c_{-k\uparrow}>\,=\,<c_{k\downarrow}c_{-k\downarrow}> \ee
The classification for spin-triplet functions $\Delta_k^{ap}=-\Delta_{-k}^{ap}$ and $\Delta_k^{p}=-\Delta_{-k}^{p}$  in the case of simple cubic lattice \cite{RKS10} inspires to look for a gap in the form with $T_{1u}$ configuration
\bea\label{iron20a}
\Delta_k^{ap} & = & \Delta^{ap}\left(\sin k_x+\sin k_y+\sin k_z \right) \nonumber \\
\\
\Delta_k^{p} & = & \Delta^{p}\left(\sin k_x+\sin k_y+\sin k_z \right), \nonumber \eea
where $ \Delta^{ap}$ and $ \Delta^{p}$ are real parameters.

For a system with Hamiltonian (\ref{nematicSc10}) and gap functions (\ref{iron20a}) the free energy, at zero temperature, as a function of the gap parameters $ \Delta^{ap}$ and  $\Delta^{p}$ has the form
\bea\label{nematicSc11}
\textit{F}_{T=0}& = & \,\frac {3}{2\kappa}\left[(\Delta^{p})^2 - (\Delta^{ap})^2\right] \\
 & + & \frac 1N \sum\limits_{k}\left[\varepsilon_k - \frac 12 E^{+}_k- \frac 12 E^{-}_k\right],\nonumber \eea
where
\be\label{nematicity12}
E^{\pm}_k=\sqrt{(\varepsilon_k)^2 + (\Delta^{a}\pm\Delta^{ap})^2 (e_k)^2},\ee
with $e_k\,=\,\sin k_x + \sin k_y + \sin k_z$, are the dispersions of the Bogoliubov quasiparticles. The gap equations are the equations for the minimum of the free energy as a function of the gap parameters
\be\label{nematicSc13}
\frac {\partial \textit{F}_{T=0}}{\partial \Delta^{p}}\,=\,0, \quad \frac {\partial \textit{F}_{T=0}}{\partial \Delta^{ap}}\ee
One can write them in the form
\bea\label{nematicSc14}
\Delta^{p} & + & \Delta^{ap}\, =\, \kappa\left(\Delta^{p}-\Delta^{ap}\right) \nonumber\\
& \times & \int\frac {d^3k}{(2\pi)^3}\frac {(e_k)^2}{\sqrt{\left(-2t\varepsilon_k-\mu\right)^2+\left(\Delta^{p}- \Delta^{ap}\right)^2(e_k)^2}} \nonumber\\
\\
\Delta^{p} & - & \Delta^{ap}\, =\, \kappa\left(\Delta^{p}+\Delta^{ap}\right) \nonumber\\
& \times & \int\frac {d^3k}{(2\pi)^3}\frac {(e_k)^2}{\sqrt{\left(-2t\varepsilon_k-\mu\right)^2+\left(\Delta^{p}+\Delta^{ap}\right)^2(e_k)^2}}. \nonumber\eea

The system of equations (\ref{nematicSc14}) has no solution with both parameters
others than zero ($\Delta^{p}\,\neq\,0,\, \Delta^{ap}\,\neq\,0$). When the coupling constant is positive ($\kappa>0$) the solution is in the form ($\Delta^{p}\,\neq\,0,\, \Delta^{ap}\,=\,0$), while for negative constants ($\kappa<0$) it is ($\Delta^{p}\,=\,0,\, \Delta^{ap}\,\neq\,0$).

\begin{figure}[!ht]
\epsfxsize=\linewidth
\epsfbox{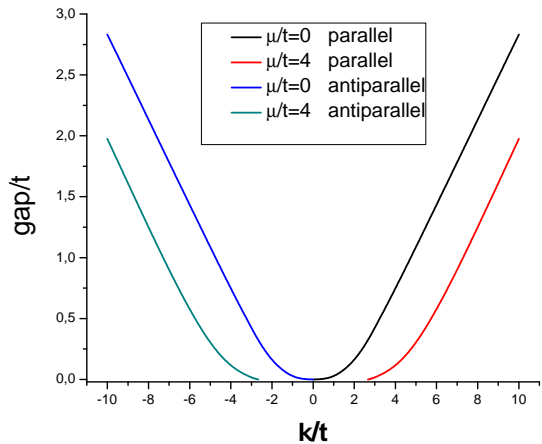}
\caption{(Color online)\, The right graphs represent the dimensionless gap $\Delta^{p}/t$ as a function of a dimensionless coupling constant $\kappa/t$ for $\mu=0$-upper graph) and $\mu/t=4$-lower one (red). The left graphs image the dimensionless gap $\Delta^{ap}/t$ as a function of a dimensionless coupling constant for $\mu=0$ (upper graph  and $\mu/t=4$ (lower one).\,}\label{fig1-gap}
\end{figure}

In both cases, the system of equations (\ref{nematicSc14}) reduces to an equation of the form
\be\label{nematicSc15}
\Delta\, =\, |\kappa|\Delta
\int\frac {d^3k}{(2\pi)^3}\frac {(e_k)^2}{\sqrt{\left(-2t\varepsilon_k-\mu\right)^2+\left(\Delta\right)^2(e_k)^2}},\ee
where $\Delta\,=\,\Delta^{p}$ when $\kappa>0$ and $\Delta\,=\,\Delta^{ap}$ if $\kappa<0$. The solutions of the equation (\ref{nematicSc15}) and respectively of the system (\ref{nematicSc14}) are depicted in Fig.(\ref{fig1-gap}). The right graphs represent the dimensionless gap $\Delta^{p}/t$ as a function of a dimensionless coupling constant $\kappa/t$ for $\mu=0$-upper graph and $\mu/t=4$-lower one. The left graphs image the dimensionless gap $\Delta^{ap}/t$ as a function of a dimensionless coupling constant for $\mu=0$-upper graph and $\mu/t=4$-lower one.

\begin{figure}[!ht]
\epsfxsize=\linewidth
\epsfbox{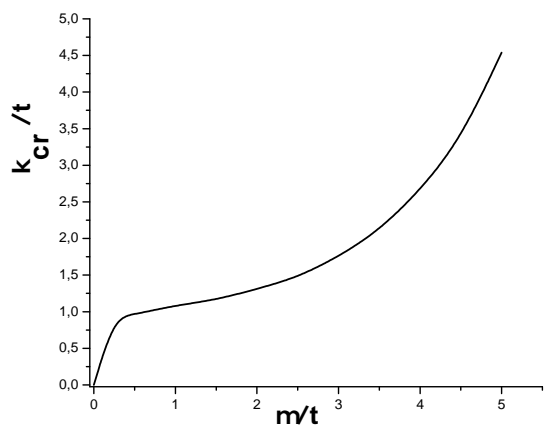}
\caption{(Color online)\,The critical constant $\kappa_{cr}/t$ as a function of the chemical potential $\mu/t$}\label{fig2-cr}
\end{figure}

The main result, evident from  \ref{fig1-gap}, is that for a system with zero chemical potential even a very small coupling can bind fermions into a bound state that leads to superconductivity. When the chemical potential is non-zero, the system possesses quantum critical transition from a normal spin-nematic phase to a phase where superconductivity coexists with spin-nematicity. One can obtain the equation for the critical value of the coupling constant $\kappa_{cr}$ from equation (\ref{nematicSc15})
\be\label{nematicSc16}
1\, =\, |\kappa_{cr}|
\int\frac {d^3k}{(2\pi)^3}\frac {(e_k)^2}{\sqrt{\left(-2t\varepsilon_k-\mu\right)^2}}.\ee
The critical value of the dimensionless constant $\kappa_{cr}/t$ as a function of the dimensionless chemical potential $\mu/t$ is depicted in Fig.(\ref{fig2-cr})

\section{Conclusion}

The appearance of unconventional superconductivity in Fe-based systems is commonly thought to
arise from a spin fluctuation pairing mechanism \cite{Johnston10,Mazin10,Graser09,Chubukov12}.
However, the unique properties of nonmagnetic $FeSe$ provides a case to test this view and gives an opportunity  for theoretical investigation of a new mechanism of superconductivity.

The present paper highlights the possibility of superconductivity induced by spin-nematic order. It is shown that in spin-nematic phase fermion-quadrupolar interaction in a spin-fermion system leads to spin triplet superconductivity with $T_{1u}$ configuration. The system is spin-nematic below critical value of the fermion-quadrupolar coupling constant and undergoes a quantum  phase transition to a phase where superconductivity coexists with spin-nematicity when the coupling constant increases. The critical value depends on the chemical potential. It is zero when the chemical potential is zero.

In the nematic but not superconducting phase the fermion-quadrupolar interaction (\ref{nematicSc7}) can be treated in a mean-field approximation to obtain a fermion system with broken rotational symmetry but invariant under the time reversal transformation. This fermion-nematic system is itself important and deserve separate consideration.

The model (\ref{nematicSc1},\ref{nematic1},\ref{nematicSc2},\ref{nematicSc6}) is discussed  in an attempt to study the nematic order and superconductivity in Fe-chalcogenide $FeSe$.
The nematic state of $FeSe$ is tunable under applied pressure \cite{FeSePressure09,FeSePressure15,FeSePressure16,FeSePressure16b,FeSePressure10,FeSePressure11}.
In the low-pressure region $0\leq p\lesssim 0.8 GPa $ the superconducting critical temperature $Tc$ increases linearly with pressure $p$ and no magnetic order is observed \cite{FeSePressure10,FeSePressure11}. This region is well described by the theory presented in the present paper.
Above $0.8 GPa$, in the intermediate pressure range $(0.8 \leqslant p \leqslant 1.2 PGa)$ the superconducting critical temperature $T_c$ is
suppressed as soon as magnetic order appears, leading to
the local minimum. For $p \gtrsim 1.2 GPa$ antiferromagnetism is fully settled, and both $T_N$ and $T_c$ increase simultaneously
with increasing pressure \cite{FeSePressure10,FeSePressure11}. To explain the transition from nematic state to antiferromagnetic order one has to consider the $\gamma$ angle in the Hamiltonian (\ref{nematic1}) as a function of the pressure. In the low-pressure region $0\leq p\lesssim 0.8 GPa $
$\gamma$ runs the interval $\frac 54 \pi< \gamma < \frac 32 \pi$ and the system is nematic. Above $0.8GPa$ the angle runs the interval $-\frac 12 \pi< \gamma < \frac 14 \pi$ and the system is antiferromagnetic. To explain the coexistence of superconductivity and antiferromagnetism one has to add an extra spin-fermion interaction term to the Hamiltonian (\ref{nematicSc1})
\be\label{nematicSc17}
H_{sf}\, = \, g \sum\limits_{i\lambda}c_{i}^{+}\tau^{\lambda} c_{i}S^{\lambda}_i,
\ee
where $\tau^{x},\tau^{y}\tau^{z}$ are Pauli matrices. This term is not important when the system is in a spin nematic state but it is paramount to describe the coexistence of the antiferromagnetism and superconductivity. These comments are inspired by the experiments reported in \cite{Imai09}.

In the present paper, we discussed three-dimensional systems but the results are qualitatively the same and for the two-dimensional ones.

\end{document}